\documentclass[11pt,a4paper]{article}

\usepackage{amsmath,amssymb,amsthm}
\usepackage{graphicx}
\usepackage{subcaption}
\usepackage{varioref}

\newtheorem{theorem}{Theorem}[section]

\newtheorem{proposition}[theorem]{Proposition}
\newtheorem{definition}[theorem]{Definition}

\newcommand{\R}{\mathbb{R}}

\newcommand{\D}{\mathcal{D}}

\newcommand{\p}{\partial}
\newcommand{\contr}{\lrcorner}

\newcommand{\cont}{\lrcorner}
\newcommand{\Span}{\operatorname{span}}
\newcommand{\Tr}{\operatorname{tr}}

\begin{document}
	
	\title{Solvable Structures for Hamiltonian Systems}

	\author{Sa\v{s}a Kre\v{s}i\'{c}--Juri\'{c}\\
		Faculty of Science\\
		University of Split\\
		Rudjera Bo\v{s}kovi\'{c}a 33\\
		21000 Split, Croatia \\
		{\tt skresic@pmfst.hr}
		\and
		Concepci\'{o}n Muriel\\
		Department of Mathematics\\
		University of C\' adiz\\
		11510 Puerto Real, Spain\\
		{\tt concepcion.muriel@uca.es}
		\and
		Adrian Ruiz\\ Department of Mathematics\\
		University of C\'{a}diz\\
		11510 Puerto Real, Spain\\
		{\tt adrian.ruiz@uca.es}}

	\date{}
	
	\maketitle
	

\begin{abstract}
	We develop a geometric framework connecting solvable structures and 
	integrable Hamiltonian systems. For a completely integrable Hamiltonian system with $n$ degrees of freedom,
	we construct a solvable structure consisting of $2n$ Hamiltonian vector fields that allows us 
	to integrate Hamilton's equations by quadratures. Explicit expressions are derived for the associated Pfaffian forms whose integration yields both the solutions of Hamilton's equations and action--angle variables. 
	This provides a novel interpretation of the Arnold--Liouville theorem in the context of solvable structures.
	We illustrate the theory by deriving explicit solutions and canonical variables of action--angle type for the open Toda lattice and the rational Calogero--Moser system. 
\end{abstract}

action--angle variables.

\section{Introduction}

Integration of differential equations by means of symmetry methods, initiated by Lie and Cartan, is one
of the most effective tools for obtaining explicit solutions of both ordinary and partial differential 
equations. For ordinary differential equations (ODEs), local symmetries can reduce the order of an equation. 
In particular, if an $n$--th order ODE admits an $n$--dimensional solvable Lie algebra of symmetries, then it can be integrated by successive quadratures. This approach, developed in Refs. \cite{Olver,Ovsiannikov1982,Stephani,BlumanAnco,Duzhin1991} provides a deep geometric explanation of integrability. 

However, the discovery of equations integrable by quadratures even in the absence of Lie point symmetries, 
such as the examples presented in \cite{artemioecu} and \cite{gonzalezgascon}, motivated the development
of new theories of integrability. Among these is the concept of a solvable structure 
introduced independently by Basarab-Horwath \cite{Basarab} and Sherring and Prince \cite{Sherring}, and further
developed in Refs. \cite{hartl1994solvable,barco2001similarity,Barco2001,Barco2002,BarcoPDE}. This framework
provides a geometric tool for integrating involutive distributions by quadratures, expressed through nested
symmetry conditions, thereby broadening the scope of symmetry--based methods for differential equations. 

In parallel, Hamiltonian mechanics gives a natural setting for integrable dynamics using symplectic
and Poisson geometry. According to the Arnold--Liouville theorem \cite{Arnold}, if a Hamiltonian system with $n$ 
degrees of freedom admits $n$ functionally independent first integrals in involution whose common
level set is compact and connected, then its motion can be described by action--angle variables in which the flow becomes linear. While
the theorem guarantees their local existence and gives a qualitative description of the orbits, explicit
construction of the action--angle variables requires identification of $n$ independent cycles on the
common level set, which is often difficult in high dimensions. Therefore, explicit integration of Hamiltonian
systems typically relies on other methods, such as the Hamilton--Jacobi equation, Lax pair representation or algebraic--geometric techniques. For a comprehensive survey of different integration techniques see \cite{Babelon}.

This paper develops an integration method for Liouville--integrable Hamiltonian systems based on solvable structures. We prove that every completely integrable Hamiltonian system with $n$ degrees of freedom admits a canonical solvable structure consisting of $2n$ Hamiltonian vector fields.
This approach leads to solutions of Hamilton's equations based on integration of $2n$ Pfaffian forms associated with the solvable structure. The last $n$ Pfaffian forms are exact and determine the action variables. The first $n$ forms become exact when restricted to the level sets of the action variables, and their integration yields the corresponding angle variables.
In this framework, Liouville integrability is reinterpreted geometrically as the integration of an involutive distribution by means of a canonical solvable structure. This viewpoint may be useful for further developments, including the study of dissipative or contact Hamiltonian systems, bi--Hamiltonian systems, and systems without a known Lax representation, thereby providing a bridge between classical integrability and a broader geometric theory of Hamiltonian systems.

The paper is organized as follows. Section 2 reviews the basic properties of solvable structures and
their role in integrating involutive distributions and ODEs by quadratures. Section 3 develops the canonical solvable structure for Hamiltonian
systems and derives explicit expressions for the associated Pfaffian forms leading to the construction
of action--angle variables. Section 4 illustrates the method by two classical examples: the open Toda lattice and the rational Calogero--Moser system. The solvable structure for the Calogero--Moser system is constructed in the physical variables, while that for the Toda lattice uses spectral variables derived from the Weyl function associated with the Lax matrix.

\section{Solvable structures for involutive distributions}
\label{Sec-2}

In this section we review the concept of solvable structures as a method for (local) integration
of involutive distributions. Throughout this section all vector fields and differential forms are assumed to be defined on an open convex domain $U\subseteq \R^m$. By $\mathfrak{X}(U)$ and $\Omega^k (U)$ we denote the $C^\infty(U)$--modules of smooth vector fields and smooth $k$--forms on $U$, respectively. A key notion in this context is that of a symmetry of a distribution. 

\begin{definition}
	Let $\D=\Span\{X_1, \ldots, X_r\}$ be a rank--$r$ smooth distribution spanned by 
	vector fields $X_i\in \mathfrak{X}(U)$, $1\leq i \leq r$. A vector field $Y\in \mathfrak{X}(U)$ is called 
	a symmetry of the distribution $\D$ if 
	\begin{enumerate}
		\item $X_1, \ldots, X_r, Y$ are pointwise linearly independent on $U$,
		\item $[Y,X_i]\in \D$ for $i=1, \ldots, r$. 
	\end{enumerate}
\end{definition}
If $Y$ is a symmetry of $\D$, then the flow $\Phi_t$ of $Y$ preserves the distribution $\D$ since 
$\Phi^\ast_t X_i \in \D$ for $i=1, \ldots, r$. 

In this paper we are primarily concerned with Frobenius integrable distributions. Although 
Frobenius integrability is equivalent to involutivity of a given distribution, for the purpose of this work it is convenient to express the integrability condition in terms of differential forms. Recall that if $\D$ is a rank--$r$ distribution on $U\subseteq \R^m$, then $\D$ can be represented as $\D=\ker \omega_1 \cap \ldots \cap \ker \omega_{m-r}$ where $\omega_1, \ldots, \omega_{m-r}$ are linearly independent one--forms on $U$ that annihilate $\D$. By the Frobenius theorem \cite{Morita}, the distribution $\D$ is integrable if and only if the Pfaffian forms $\omega_i$ satisfy the integrability condition
\begin{equation}\label{int-cond}
	d\omega_k = \sum_{j=1}^{m-r} \theta_{kj} \wedge \omega_j \quad \text{for some}\quad 
	\theta_{kj}\in \Omega^1 (U).
\end{equation}
The integrability condition \eqref{int-cond} implies that the ideal $I(\omega_1, \ldots, \omega_{m-r})$ 
generated by $\omega_i\in \Omega^1 (U)$, $1\leq i \leq m-r$, is a differential ideal in the algebra
$\Omega (U) = \bigoplus_{k=1}^{m-r}\, \Omega^k (U)$ since $dI \subseteq I$. Under certain conditions,
the integral manifold of $\D$ can be found by means of a solvable structure associated with $\D$. 

\begin{definition}
	Let $\D=\Span\{X_1, \ldots, X_r\}$ be a rank--$r$ involutive distribution on $U\subseteq \R^m$. A solvable structure for $\D$ is a sequence of vector fields $(X_1, \ldots, X_r, Y_1, \ldots, Y_{m-r})$ such that 
	\begin{enumerate}
		\item $X_1, \ldots, X_r, Y_1, \ldots, Y_{m-r}$ are pointwise linearly independent on $U$, 
		\item $Y_1$ is a symmetry of $\D$,
		\item $Y_k$ is a symmetry of $\D_{k-1} = \D \oplus \Span \{Y_1, \ldots, Y_{k-1}\}$, $k=2, \ldots, m-r$. 
	\end{enumerate}
\end{definition}
In general, the vector fields $Y_1, \ldots, Y_{m-r}$ need not form a solvable Lie algebra. Thus, solvable structures are more flexible and apply beyond the classical solvable symmetry algebra approach to integration of distributions. The following theorem gives a necessary and sufficient condition for an involutive distribution $\D$ to admit a solvable structure: its associated Pfaffian system must be reducible to a triangular form \cite{hartl1994solvable}.

\begin{theorem}\label{Tm-2.1}
	Assume that $\D = \ker \omega_1 \cap \ldots \cap \ker \omega_{m-r}$ is an involutive distribution
	of rank $r$ on a convex domain $U\subseteq \R^m$. Then $\D$ admits a solvable structure if and only if
	\begin{equation}\label{SS-condition}
		d\omega_k = \sum_{j=k+1}^{m-r} \theta_{kj}\wedge \omega_j, \quad k=1, \ldots, m-r-1
	\end{equation}
	for some $\theta_{kj}\in \Omega^1 (U)$, with $d\omega_{m-r}=0$. 
\end{theorem}
Equivalently, $\D$ admits a solvable structure if and only if $d\omega_k \in I(\omega_{k+1}, \ldots, \omega_{m-r})$. If $(X_1, \ldots, X_r,Y_1,\ldots, Y_{m-r})$ is a solvable structure for the distribution $\D=\Span \{X_1, \ldots, X_r\}$, then the Pfaffian forms in Theorem \ref{Tm-2.1} can be represented by 
\begin{equation}
	\omega_k = \frac{1}{\lambda} Y_{m-r} \contr \ldots \contr \hat Y_k \contr \ldots \contr Y_1 \contr X_r \contr \ldots \contr X_1 \contr \tau,
\end{equation}
where $\tau$ is the volume form on $\R^m$, $\hat Y_k$ denotes omission of the vector field $Y_k$ and 
$\lambda = Y_{m-r} \contr \ldots \contr Y_1 \contr X_r \contr \ldots \contr X_1 \contr \tau$ is a smooth function on $U$.

Using Theorem \ref{Tm-2.1}, integral manifolds of $\D=\ker \omega_1 \cap \ldots \cap \ker \omega_{m-r}$ can be locally constructed by successive quadratures as follows. Since $d\omega_{m-r}=0$, the Poincar\'{e} lemma implies that there exists $I_{m-r}\in C^\infty (U)$ such that $\omega_{m-r}=dI_{m-r}$. For a fixed constant $c_{m-r}\in \operatorname{Im}(I_{m-r})$, define the submanifold $M_{m-r}=\{x\in U\mid I_{m-r}(x)=c_{m-r}\}$. 
Condition \eqref{SS-condition} implies that $d\omega_{m-r-1}\in I(\omega_{m-r})$, hence the restriction of $\omega_{m-r-1}$ to $M_{m-r}$ is closed. Therefore, there exists $I_{m-r-1}\in C^\infty (M_{m-r})$ such that $\omega_{m-r-1}\vert_{M_{m-r}} = dI_{m-r-1}$. Iterating this argument yields the integral manifolds of $\D$ locally given by the common level sets	
\begin{equation}
M_1=\big\{x\in U\mid I_1(x)=c_1,\, \ldots,\,  I_{m-r}(x)=c_{m-r}\big\}.
\end{equation}

The above result can be used to integrate a system of ordinary differential equations (ODEs) by applying the procedure to the rank--one distribution $\D=\Span\{A\}$, where $A$ is the vector field on a jet space representing the system. In this case, if $(A,Y_1,\ldots,Y_{m-1})$ is a solvable structure for $\D$, then the Pfaffian forms that annihilate $\D$ are given by
\begin{equation}\label{Pfaff}
	\omega_k=\frac{1}{\lambda}\, Y_{m-1}\cont \ldots \cont \hat{Y}_k \cont \ldots \cont Y_1 \cont A \cont \tau, \quad k=1, \ldots, m-1.
\end{equation}
Successive integration of the forms \eqref{Pfaff} as described above yields
integral curves of $A$, which represent solutions of a given system of ODEs (for more details see \cite{Basarab,hartl1994solvable,Barco2001,warner}).

Because of the symplectic nature of Hamiltonian systems, we will show in the next section that a canonical solvable structure can be defined for a completely integrable Hamiltonian system. The corresponding Pfaffian forms \eqref{Pfaff} can then be obtained explicitly in terms of the Hamiltonian and its first integrals in involution, leading to explicit solutions and the action--angle variables of the system.

\section{Solvable structures for Hamilton's equations}

In Hamiltonian mechanics, a system with $n$ degrees of freedom is described by points $(q,p)$ on a $2n$--dimensional symplectic manifold $(M,\omega)$. Here, $q=(q_1, \ldots, q_n)$ represent generalized coordinates and $p=
(p_1, \ldots, p_n)$ the conjugate momenta of the system. The nondegeneracy of the symplectic form $\omega$ implies
that every Hamiltonian $H\in C^\infty (M)$ determines a unique Hamiltonian vector field $X_H\in \mathfrak{X}(M)$ by
\begin{equation}\label{X_H}
	X_H \cont\, \omega = dH .
\end{equation}
The integral curves of $X_H$ describe the time evolution of the system. By Darboux's theorem, in a neighbourhood of every point of $M$ there exist local coordinates $(q,p)$ in which $\omega$ has the canonical form
\begin{equation}
\omega = \sum_{i=1}^n dq_i \wedge dp_i.
\end{equation}
In these coordinates,
\begin{equation}
	X_H=\sum_{i=1}^n
	\frac{\partial H}{\partial p_i}\frac{\partial}{\partial q_i}
	-
	\frac{\partial H}{\partial q_i}\frac{\partial}{\partial p_i},
\end{equation}
and the integral curves of $X_H$ satisfy Hamilton's equations
\begin{equation}\label{can_Ham}
	\dot q_i=\frac{\partial H}{\partial p_i},\qquad
	\dot p_i=-\frac{\partial H}{\partial q_i},\qquad i=1, \ldots,n .
\end{equation}
The flow $\varphi_t$ of $X_H$ preserves both the symplectic form $\omega$ and the Hamiltonian $H$, that is
$\varphi_t^{\ast}\omega=\omega$ and $\varphi_t^\ast H = H$. 

In applications of solvable structures to Hamilton's equations, an important role is played by the Poisson bracket of two functions $F,G\in C^\infty (M)$ defined by
\begin{equation}\label{Poisson}
	\{F,G\}=\omega(X_F,X_G).
\end{equation}
In Darboux coordinates, the Poisson bracket takes the form
\begin{equation}\label{PB-9}
	\{F,G\}=\sum_{i=1}^n 
	\frac{\partial F}{\partial q_i}\frac{\partial G}{\partial p_i}
	-
	\frac{\partial G}{\partial q_i}\frac{\partial F}{\partial p_i}.
\end{equation}
The evolution of any observable $F\in C^\infty(M)$ along the flow $\varphi_t$ of the vector field $X_H$ is given by
\begin{equation}\label{dot_F}
	\frac{d}{dt} (F\circ \varphi_t) = \{F,H\}\circ \varphi_t,
\end{equation}
or simply $\dot F = \{F,H\}$. Thus, if $F$ and $H$ are in involution, that is $\{F,H\}=0$, then $F$ is a first integral of $X_H$. Moreover, 
\begin{equation}
[X_F,X_H]=-X_{\{F,H\}}=0,
\end{equation}
so $X_F$ is a symmetry of the vector field $X_H$. We shall be concerned with solvable structures associated with completely integrable Hamiltonian systems. 
\begin{definition}
Let $(M,\omega)$ be a $2n$--dimensional symplectic manifold and let $H\in C^\infty (M)$. The Hamiltonian vector field $X_H$ is said to be completely integrable, or Liouville integrable, on an open subset $U\subseteq M$ if there exist functions $F_1, \ldots, F_n\in C^\infty (U)$ such that 
\begin{enumerate}
\item $dF_1 \wedge \ldots \wedge dF_n\neq 0$ on $U$,
\item $\{F_i,F_j\}=0$ and $\{F_i,H\}=0$, $i,j=1, \ldots, n$. 
\end{enumerate}
\end{definition}
\noindent Conditions 1 and 2 imply that $F_1, \ldots ,F_n$ are functionally independent first integrals of $X_H$ in involution.
Under additional compactness assumptions, complete integrability of $X_H$ leads to action--angle variables that linearize the motion. Define $F=(F_1, \ldots, F_n)\colon U \to \R^n$ and let $M_c=F^{-1}(c)$ be the common level set of $F_1, \ldots, F_n$. Since $dF_1, \ldots, dF_n$ are linearly independent on $U$, each $M_c$ is a regular $n$--dimensional submanifold invariant under the flow of $X_H$. If $M_c$ is compact and connected, then by the Arnold--Liouville theorem \cite{Arnold}, $M_c$ is diffeomorphic to an $n$--torus $\mathbb{T}^n$. Moreover, in a neighbourhood of $M_c$ there exist action--angle coordinates $(Q,P)=(Q_1,\ldots, Q_n,P_1, \ldots ,P_n)$ in which $H=H(P)$ and 
\begin{equation}
        \dot P_i=0,\qquad
        \dot Q_i=\frac{\partial H}{\partial P_i},
        \qquad i=1, \ldots,n.
\end{equation}
Thus, on each invariant torus, $P_i=\mathrm{const.}$ and the dynamics reduces to
the linear flow
\begin{equation}
        Q_i(t)=\frac{\p H}{\p P_i}t+Q_i(0).
\end{equation}
While the Arnold--Liouville theorem proves existence of the action--angle variables, the main difficulty in applying the theorem explicitly is the construction of such variables. In the following section we develop an approach based on solvable structures for the distribution $\D=\Span \{\p_t+X_H\}$ on the extended phase space $\R \times M$, which yields both explicit solutions and the corresponding action--angle variables.


\subsection{Construction of the canonical solvable structure}
\label{Sect.-3.1}

In this section we assume that $X_H$ is a completely integrable Hamiltonian vector field on a symplectic manifold $(M,\omega)$, $\dim M = 2n$, with independent first integrals $F_1, \ldots, F_n$ in involution. Our objective is to construct a solvable structure for the rank--one distribution $\D = \Span \{A\}$ generated by the dynamical vector field $A=\p_t + X_H$ on $\R \times M$. We look for a solvable structure of the form
\begin{equation}\label{3.10}
	(A, X_{F_1}, \ldots, X_{F_n}, X_{G_1}, \ldots, X_{G_n})
\end{equation}
where $G_i$ are smooth functions defined on a suitable open subset of $M$. 
Here, $X_{F_1}$ is a symmetry of $\D=\Span \{A\}$, and $X_{F_k}$ is a symmetry of the rank-$k$ involutive distribution 
$\D_{k-1}=\Span \{A, X_{F_1}, \ldots, X_{F_{k-1}}\}$. We remark that the partial solvable structure $(A, X_{F_1}, \ldots, X_{F_n})$ is not sufficient for the construction of the action--angle variables of the system. In order to complete it, we add independent Hamiltonian vector fields $X_{G_1}, \ldots, X_{G_n}$ such that 
\begin{enumerate}
	\item $X_{G_1}$ is a symmetry of $\D_n=\Span\{A,X_{F_1}, \ldots, X_{F_n}\}$,
	\item $X_{G_k}$ is a symmetry of $\D_n \oplus \Span\{X_{G_1}, \ldots, X_{G_{k-1}}\}, \quad k=2, \ldots, n$.
\end{enumerate}
While the integrals $F_k$ yield the action variables, we will show that the functions $G_k$ give rise to the angle variables. Thus, $G_k$ play the role of the conjugate variables to $F_k$. In particular, we prove that there exist functions $G_k$ such that
\begin{align}
	[X_{G_i},X_{G_j}] &= 0,  \label{3.12}\\
	[X_{G_i},X_{F_j}] &= \sum_{k=1}^n f_{ij}^k\, X_{F_k}, \label{3.13} \\
	[X_{G_i},A] &= \sum_{k=1}^n h_{ik}\, X_{F_k}. \label{3.14}
\end{align}
for some smooth functions $f_{ij}^k$ and $h_{ik}$ defined on an open subset of $M$. First, we prove the claim with $f_{ij}^k=0$, and later we extend the proof to the more general Lie bracket \eqref{3.13}.

\begin{theorem}\label{tm-3.2}
Let $(M,\omega)$ be a $2n$--dimensional symplectic manifold and let $X_H$ be a completely integrable Hamiltonian vector field on $M$ with functionally independent first integrals $F_1, \ldots, F_n\in C^\infty (M)$ in involution. If the common level set $M_c=F^{-1}(c)$ is compact and connected, then in a neighbourhood of $M_c$ there exist locally defined functions $G_1, \ldots, G_n$ such that $\{G_i,G_j\}=0$, $\{G_i,F_j\}=\delta_{ij}$. Consequently, on the extended phase space $\R \times M$, the sequence of vector fields \eqref{3.10} forms a local solvable structure for the distribution $\D=\Span \{A\}$ such that
\begin{equation}
[X_{G_i},X_{G_j}]=0, \quad [X_{G_i},X_{F_j}]=0, \quad [X_{G_i},A]\in \Span \{X_{F_1}, \ldots, X_{F_n}\}.
\end{equation}
\end{theorem}
	
\proof By the Arnold--Liouville theorem, there exists a neighbourhood $V$ of $M_c$ and action--angle variables $(Q,P)=(Q_1, \ldots, Q_n, P_1, \ldots, P_n)$ such that the symplectic form is given by $\omega=\sum_{i=1}^n dQ_i \wedge dP_i$. In these coordinates, the Hamiltonian $H$ and first integrals $F_i$ depend only on the action variables $P_i$, that is $H=H(P)$ and $F_i=F_i(P)$. Since $F_1, \ldots, F_n$ are functionally independent, the Jacobian matrix $DF(P)=\big[\frac{\p F_i}{\p P_j}\big]$ is nonsingular. Hence, we can locally invert the map $(P_1, \ldots, P_n)\mapsto (F_1, \ldots, F_n)$ and write $P_i = P_i (F_1, \ldots , F_n)$, $i=1,\ldots, n$. 

The angle coordinates $Q_i$ are defined modulo periods, hence, if necessary, we restrict to a local neighbourhood in $V$ and choose local representatives of $Q_1, \ldots, Q_n$. Define the generating function
\begin{equation}
	S(F,Q) = \sum_{i=1}^n P_i(F) Q_i
\end{equation}
and set
\begin{equation}\label{20-A}
	G_i = \frac{\p S}{\p F_i} = \sum_{k=1}^n \frac{\p P_k}{\p F_i} \, Q_k.
\end{equation}
Next, we verify the Poisson brackets $\{G_i,F_j\}$ and $\{G_i,G_j\}$. Since $\omega=\sum_{i=1}^n dQ_i \wedge dP_i$, the Poisson bracket in the action--angle coordinates is given by the canonical form \eqref{PB-9}. We obtain
\begin{equation}
\{G_i,F_j\} = \sum_{k=1}^n \frac{\p G_i}{\p Q_k} \frac{\p F_j}{\p P_k} = \sum_{k=1}^n \frac{\p P_k}{\p F_i} \frac{\p F_j}{\p P_k} = \delta_{ij}.
\end{equation}
Furthermore,
\begin{align}
	\{G_i,G_j\} &= \sum_{k=1}^n \frac{\p G_i}{\p Q_k} \frac{\p G_j}{\p P_k} - \frac{\p G_j}{\p Q_k} 
	\frac{\p G_i}{\p P_k} \notag \\
	&= \sum_{l=1}^n \left[\sum_{k=1}^n \frac{\p P_k}{\p F_i} \frac{\p}{\p P_k} \Big(\frac{\p P_l}{\p F_j}\Big)
	-\frac{\p P_k}{\p F_j}\frac{\p}{\p P_k}\Big(\frac{\p P_l}{\p F_i}\Big)\right]  Q_l \notag \\
	&= \sum_{l=1}^n \left[\frac{\p^2 P_l}{\p F_i \p F_j}-\frac{\p^2 P_l}{\p F_j \p F_i}\right] Q_l = 0.
\end{align}
It follows that $[X_{G_i},X_{G_j}]=0$ and $[X_{G_i},X_{F_j}]=0$. It remains to compute the bracket $[X_{G_i},A]$. We have 
\begin{equation}
[X_{G_i},A]=-X_{\{G_i,H\}}=-\left(\sum_{k=1}^n \frac{\p K_i}{\p P_k} \frac{\p}{\p Q_k}-\frac{\p K_i}{\p Q_k}
\frac{\p}{\p P_k}\right)
\end{equation}
where the function $K_i=\{G_i,H\}$ is given by
\begin{equation}
K_i = \sum_{k=1}^n \frac{\p P_k}{\p F_i} \frac{\p H}{\p P_k}.
\end{equation}
Since $\frac{\p P_k}{\p F_i}$ depends only on the action variables $P_i$, it follows that $K=K(P)$ and consequently
\begin{equation}\label{25-C}
[X_{G_i},A]=-\sum_{k=1}^n \frac{\p K_i}{\p P_k} \frac{\p}{\p Q_k}.
\end{equation}
The vector fields
\begin{equation}\label{26-C}
X_{F_i} = \sum_{k=1}^n \frac{\p F_i}{\p P_k} \frac{\p}{\p Q_k}, \quad i=1, \ldots, n,
\end{equation}
form a basis of the tangent space to the invariant tori, hence $[X_{G_i},A]$ can be expressed as a linear combination of $X_{F_1}, \ldots, X_{F_n}$. Therefore, Eq. \eqref{3.14} holds where $h_{ij}$ are the entries of the matrix $-DK(P) \, DF(P)^{-1}$ with $DK(P)= \big[\frac{\p K_i}{\p P_j}\big]$. 

Finally, from the Poisson bracket relations $\{F_i,F_j\}=0$, $\{G_i,G_j\}=0$ and $\{G_i,F_j\}=\delta_{ij}$ it follows that the vector fields $A,X_{F_1}, \ldots, X_{F_n}, X_{G_1}, \ldots, X_{G_n}$ are linearly independent on $\R \times V$. Therefore, they form a local solvable structure for $\D = \Span \{A\}$. \qed
\bigskip

We refer to the solvable structure in Theorem \ref{tm-3.2} as the canonical solvable structure. We emphasize that the functions $G_i$ are not necessarily globally defined on all of $M$. Using Eq. \eqref{20-A}, they can be locally expressed in terms of the first integrals as
\begin{equation}\label{22-A}
G=\big(DF(P)^{-1}\big)^T Q
\end{equation}
where $G=[G_1 \ldots G_n]^T$ and $Q=[Q_1 \ldots Q_n]^T$. 

The compactness and connectedness assumptions in Theorem \ref{tm-3.2} are used in order to apply the Arnold--Liouville theorem and interpret $G_i$ and $F_i$ as action--angle variables on a neighbourhood of a regular invariant torus. These assumptions are not needed for purely local construction of a solvable structure. Indeed, if $F_1, \ldots, F_n$ are functionally independent and in involution on an open subset $U\subset M$, then by the Carath\'{e}odory--Jacobi--Lie theorem in a neighbourhood of every point of $U$ there exist functions $G_1, \ldots, G_n$ such that $\{G_i,G_j\}=0$ and $\{G_i,F_j\}=\delta_{ij}$. This explains why the above construction of the solvable structure also applies to noncompact integrable systems as the ones discussed in Section 4. 

We next extend the above construction to a more general solvable structure with $f_{ij}^k\neq 0$. 

\begin{theorem}\label{tm-3.3}
Let $(M,\omega)$ be a $2n$--dimensional symplectic manifold, and let $X_H$ be a completely integrable Hamiltonian vector field with first integrals $F_1, \ldots, F_n$ in involution. Let $U\subset M$ be an open subset on which $F_1, \ldots, F_n$ are functionally independent. Assume that on $U$ there exist functions $G_1, \ldots, G_n$ satisfying $\{G_i,G_j\}=0$ and $\{G_i,F_j\}=\delta_{ij}$. Define
\begin{equation}\label{29-A}
\tilde G_i =\sum_{j=1}^n K_{ij}(F)\, G_j, \quad i=1, \ldots, n,
\end{equation}
where $K=\big[K_{ij}\big]$ is a nonsingular matrix of smooth functions $F=(F_1, \ldots, F_n)$ such that 
\begin{equation}\label{30-A}
	\sum_{l=1}^n K_{il} \frac{\p K_{jk}}{\p F_l} - K_{jl} \frac{\p K_{ik}}{\p F_l} = 0, \quad 
		i,j,k=1, \ldots n.
\end{equation}
Then $(A,X_{F_1}, \ldots, X_{F_n}, X_{\tilde G_1}, \ldots, X_{\tilde G_n})$, $A=\p_t+X_H$, is a local solvable structure for the distribution $\D = \Span \{A\}$ such that 
\begin{equation}
	[X_{\tilde G_i},X_{\tilde G_j}]=0, \quad [X_{\tilde G_i},X_{F_j}]\in \Span\{X_{F_1}, \ldots, X_{F_n}\}, \quad [X_{\tilde G_i},A]\in \Span\{X_{F_1}, \ldots, X_{F_n}\}.
\end{equation}
\end{theorem}

\proof We first prove that the functions $\tilde G_i$ satisfy the Poisson bracket
\begin{equation}\label{31-A}
	\{\tilde G_i, \tilde G_j\}=0, \quad \{\tilde G_i,F_j\}=K_{ij}(F).
\end{equation}
The Poisson bracket of $\tilde G_i$ and $F_j$ yields
\begin{equation}\label{33-C}
	\{\tilde G_i,F_j\} = \sum_{k=1}^n K_{ik} \{G_k,F_j\} = K_{ij}.
\end{equation}
Next consider the Poisson bracket
\begin{align}
	\{\tilde G_i,\tilde G_j\}
	&= \sum_{k=1}^n \sum_{l=1}^n \Big(K_{ik} \{G_k, K_{jl}\} G_l + K_{jl} \{K_{ik},G_l\} G_k\Big) \notag \\
	&=\sum_{k=1}^n \left(\sum_{l=1}^n K_{il} \frac{\p K_{jk}}{\p F_l} - K_{jl} \frac{\p K_{ik}}{\p F_l}\right) G_k.
\end{align}
Hence, $\{\tilde G_i,\tilde G_j\}=0$ if and only if the functions $K_{ij}$ satisfy the system of equations \eqref{30-A} for all $i,j,k=1, \ldots, n$. Furthermore, Eq. \eqref{33-C} implies that 
\begin{equation}
	[X_{\tilde G_i},X_{F_j}] = -X_{\{\tilde G_i,F_j\}} = 
	-\sum_{k=1}^n \frac{\p K_{ij}}{\p F_k} X_{F_k} \in \Span \{X_{F_1}, \ldots, X_{F_n}\}.
\end{equation}
Since $F_1, \ldots, F_n$ are functionally independent on $U$, the map $F=(F_1,\ldots, F_n)\colon U\to 
\R^n$ is a submersion. The vector fields $X_{F_1}, \ldots X_{F_n}$ span the tangent spaces to the level sets $F^{-1}(c)$. Since $\{H,F_i\}=0$ for $i=1,\ldots, n$, the function $H$ is constant on the level sets $F^{-1}(c)$. Therefore, after possibly replacing $U$ by a smaller open subset, there exists a function $h$ such that $H=h\circ F$ on $U$. By abuse of notation, we may write $H=H(F_1,\ldots, F_n)$. This yields
\begin{equation}
\{\tilde G_i,H\} = \sum_{j=1}^n K_{ij}\, \{G_j,H\} =\sum_{j=1}^n K_{ij}\, \frac{\p H}{\p F_j}
\end{equation}
implying that 
\begin{equation}
[X_{\tilde G_i},A] = -X_{\{\tilde G_i,H\}} \in \Span \{X_{F_1}, \ldots, X_{F_n}\}.
\end{equation}
This proves that $(A,X_{F_1}, \ldots, X_{F_n}, X_{\tilde G_1}, \ldots, X_{\tilde G_n})$ is a local solvable structure for $\D=\Span \{A\}$ of the form \eqref{3.12}--\eqref{3.14} with $f_{ij}^k = - \frac{\p K_{ij}}{\p F_k}$. 
\qed
\bigskip

We note that the Poisson bracket \eqref{31-A} together with $\{F_i,F_j\}=0$ must satisfy the Jacobi identity. The only nontrivial identity comes from the Jacobiator
\begin{equation}
\operatorname{Jac}(\tilde G_i, \tilde G_j, F_k)=\{\{\tilde G_i,\tilde G_j\},F_k\}+\{\{\tilde G_j,F_k\},\tilde G_i\}+\{\{F_k,\tilde G_i\},\tilde G_j\}
\end{equation}
which vanishes precisely when the condition \eqref{30-A} holds. This condition has the following geometric interpretation. Define the vector fields 
\begin{equation}\label{36-A}
Y_i = \sum_{j=1}^n K_{ij} \frac{\p}{\p F_j},  \quad i=1, \ldots, n. 
\end{equation}
Then the functions $K_{ij}$ satisfy Eq. \eqref{30-A} if and only if 
\begin{equation}\label{37-AA}
[Y_i,Y_j] = 0, \quad i,j=1, \ldots, n.
\end{equation}
Therefore, the Poisson commutativity of $\tilde G_1, \ldots, \tilde G_n$ is equivalent to the condition that the vector fields $Y_1, \ldots, Y_n$ form a commuting frame. Moreover, if $\det [K_{ij}]\neq 0$, then 
$Y_1, \ldots, Y_n$ are linearly independent. Hence, by the local straightening theorem, there exist coordinates $(u_1, \ldots, u_n)$ such that $Y_i = \frac{\p}{\p u_i}$. In these coordinates $Y_i(F_j)=\frac{\p F_j}{\p u_i}$. Since, on the other hand, $Y_i(F_j) = K_{ij}$, it follows that $K_{ij}=\frac{\p F_j}{\p u_i}$ is a solution of the PDE system \eqref{30-A}. 

\begin{proposition}
Let $\tilde G_i$, $i=1,\ldots, n$ be the functions satisfying the assumptions of Theorem \ref{tm-3.3}. Then the Poisson bracket \eqref{31-A} is invariant under the gauge transformation 
\begin{equation}
\tilde G_i \mapsto \tilde G_i + Y_i(\Phi)
\end{equation}
where $Y_1, \ldots, Y_n$ is a commuting frame defined by \eqref{36-A} and $\Phi=\Phi(F)$ is a smooth function.
\end{proposition}

\proof Consider the transformation of the form $\tilde G_i \mapsto \tilde G_i + g_i(F)$ for some smooth function $g_i(F)$. Then Eq. \eqref{31-A} implies
\begin{equation}
\{ \tilde G_i + g_i(F),F_j\}=K_{ij}(F).
\end{equation}
The Poisson bracket for $\tilde G_i+g_i$ and $\tilde G_j +g_j$ yields
\begin{equation}
\{\tilde G_i + g_i,\tilde G_j + g_j\} = \sum_{k=1}^n \left(K_{ik}\frac{\p g_j}{\p F_k}-K_{jk} \frac{\p g_i}{\p F_k}\right)=Y_i(g_j)-Y_j(g_i).
\end{equation}
If we define $g_i=Y_i(\Phi)$ for some smooth potential $\Phi=\Phi(F)$, then it follows from Eq. \eqref{37-AA} that
\begin{equation}
\{\tilde G_i+g_i,\tilde G_j+g_j\} = [Y_i,Y_j](\Phi)=0.
\end{equation}
Hence, the Poisson bracket \eqref{31-A} is invariant under the gauge transformation $\tilde G_i \mapsto \tilde G_i + Y_i(\Phi)$. \qed


\subsection{Pfaffian forms and construction of the action--angle variables}

In this section, we derive explicit formulas for the $2n$ Pfaffian forms associated with the solvable structure developed in Section 3.1, specializing to the action--angle coordinates $(Q,P)$ provided by the Arnold--Liouville theorem. In this setting, the conjugate variables $G_i$ are chosen to coincide with the angle variables $Q_i$, and successive integration of the Pfaffian forms described in Section 2 yields a quadrature--based method for recovery of the action--angle variables. Integration of the last $n$ forms enforces conservation of the action variables $P_i$, while integration of the remaining $n$ forms determines linear time evolution of the angle variables $Q_i$. 

In view of Eqs. \eqref{29-A} and \eqref{22-A}, the functions $\tilde G_i$ can be written in matrix form as
\begin{equation}
\tilde G = K(F) \big(DF(P)^{-1}\big)^T Q
\end{equation}
where $K(F)$ is the $n\times n$ matrix with elements $K_{ij}(F)$. Choosing $K(F)=DF(P)^T$, i.e. $K_{ij}=\frac{\p F_j}{\p P_i}$, we obtain $\tilde G = Q$. This choice is admissible because the
vector fields $Y_i = \sum_{j=1}^n K_{ij}\frac{\p}{\p F_j}$ commute, which is equivalent to the compatibility condition \eqref{30-A} for $K_{ij}$. 
To simplify the notation, we henceforth denote $\tilde G$ by $G$. With this choice for $G=Q$, we obtain the solvable structure \eqref{3.12}--\eqref{3.14} with 
\begin{equation}
f_{ij}^k = -\frac{\p}{\p F_k}\Big(\frac{\p F_j}{\p P_i}\Big)
\end{equation}
where the functions $G_i$ are precisely the angle variables $Q_i$. Accordingly,
 the vector fields in 
\eqref{3.12}--\eqref{3.14} are given by
\begin{equation}
	A =\frac{\p}{\p t}+ \sum_{k=1}^n \frac{\p H}{\p P_k} \frac{\p}{\p Q_k}, \quad X_{F_i} = \sum_{k=1}^n \frac{\p F_i}{\p P_k}\frac{\p}{\p Q_k}, \quad X_{G_i} = -\frac{\p}{\p P_i}.
\end{equation}
Let $\tau$ denote the volume form on $\R^{2n+1}$,
\begin{equation}
	\tau = dt \wedge dQ_1 \wedge \ldots \wedge dQ_n \wedge dP_1 \wedge \ldots \wedge dP_n,
\end{equation}
and define the function $\lambda$ by
\begin{equation}
	\lambda = X_{G_n} \cont \ldots \cont X_{G_1} \cont X_{F_n}\cont \ldots \cont X_{F_1} \cont A\cont \tau.
\end{equation}
Following the general construction in Eq. \eqref{Pfaff}, we introduce the $2n$ Pfaffian forms
\begin{align}
	\omega_k &= \frac{1}{\lambda} X_{G_n} \cont \ldots X_{G_1} \cont X_{F_n} \cont \ldots \cont \hat X_{F_k}\cont \ldots \cont X_{F_1}\cont A \cont \tau,  \label{3.40} \\
	\omega_{n+k} &= \frac{1}{\lambda} X_{G_n} \cont \ldots \cont \hat X_{G_k} \cont \ldots \cont X_{G_1} \cont X_{F_n} \cont \ldots \cont X_{F_1} \cont A \cont \tau, \quad 1\leq k \leq n, \label{3.41}
\end{align}
where $\hat X_{F_k}$ and $\hat X_{G_k}$ denote omission of the corresponding vector fields. 
A direct computation yields
\begin{equation}
	\lambda = (-1)^n \det DF (P).
\end{equation}
Since the differentials $dF_1, dF_2, \ldots,dF_n$ are linearly independent on $U\subset M$ (cf. Theorem \ref{tm-3.3}), it follows that $\lambda \neq 0$ on $U$. After a straightforward algebraic computation using \eqref{3.40}--\eqref{3.41}, the Pfaffian forms decompose naturally into $n$ exact differentials of the action variables and $n$ forms involving the angle variables. Define $f_k=(H,F_1, \ldots, \hat F_k, \ldots, F_n)\colon U\to \R^n$ where $\hat F_k$ denotes the omission of $F_k$, and let $DF(P)_{j,k}$ denote the Jacobian matrix of $F=(F_1,\ldots, F_n)\colon U\to \R^n$ with $j$--th row and $k$--th column deleted. Then the Pfaffian forms are given by
\begin{align}
\omega_{n+k} &= (-1)^{n+k+1} dP_k, \label{3.47} \\[0.2cm]
\omega_k &=\frac{(-1)^n}{\lambda} \Big(\det Df_k (P)\, dt + \sum_{j=1}^n (-1)^j \det DF(P)_{j,k} \, dQ_j\Big), \qquad 1\leq k\leq n.  \label{3.48}
\end{align}
Let $\gamma$ be an integral curve of the vector fields $A=\p_t + X_H$. According to the integration procedure outlined in Section \ref{Sec-2}, the Pfaffian forms $\omega_i$ vanish on $\gamma$. 
Since the forms \eqref{3.47} are exact, it follows that $P_k$ are constant along
$\gamma$. Define the submanifold
\begin{equation}
	N_c=\Big\{(t,Q,P)\in \R^{2n+1} \mid P_1=c_1,\, P_2=c_2,\, \ldots,\, P_n=c_n\Big\}
\end{equation}
for some $c_i\in \R$. The forms \eqref{3.48} become exact when restricted to $N_c$, that is $\omega_k \vert_{N_c}=dI_k$ where
\begin{equation}\label{3.50}
I_k = \frac{(-1)^k}{\lambda} \Big( \det Df_k(P)\, t + \sum_{j=1}^n (-1)^j \det DF(P)_{j,k}\, Q_j\Big), \qquad 1\leq k\leq n.
\end{equation}
Hence, $P_k$ and $I_k$ are first integrals of the vector field $A$. 

Next, we show that along the integral curve $\gamma$, the angle variables $Q_k$ evolve linearly in time in accordance with the Arnold--Liouville theorem. The variables $Q_k$ satisfy the system of equations
\begin{equation}\label{3.51}
	\sum_{j=1}^n \Delta_{ij}\, Q_j = B_i, \quad i=1, \ldots, n.
\end{equation}
where
\begin{equation}\label{3.49-B}
\Delta_{ij} = (-1)^{j+1} \det DF(P)_{j,i}\, , \qquad B_i = \det Df_i(P)\, t + C_i
\end{equation}
and $C_i=(-1)^{n+1} \lambda I_i$. After some algebraic computation, one finds that the determinant of the matrix $\Delta=[\Delta_{ij}]$ is given by
\begin{equation}
\det (\Delta) = (-1)^{\frac{n(n-1)}{2}} \lambda^{n-1}.
\end{equation}
Since $\lambda\neq 0$ on $U$, the system \eqref{3.51} has a unique solution given by Cramer's rule
\begin{equation}\label{3.54}
	Q_j = \frac{\det (\Delta_j)}{\det (\Delta)}
\end{equation}
where the matrix $\Delta_j$ is obtained by replacing the $j$--th column in $\Delta$ by $[B_1 \ldots B_n]^T$. 
In view of \eqref{3.49-B}, the determinant of $\Delta_j$ can be decomposed as $\det (\Delta_j)=
\det(\Delta_j^{(1)})+\det(\Delta_j^{(2)})$ where
\begin{equation}\label{3.56}
	\det(\Delta_j^{(1)}) = \begin{vmatrix}
		\Delta_{11} & \ldots & \Delta_{1,j-1} & \det Df_1 (P) & \Delta_{1,j+1} & \ldots & \Delta_{1n} \\[0.2cm]
		\Delta_{21} & \ldots & \Delta_{2,j-1} & \det Df_2 (P) & \Delta_{2,j+1} & \ldots & \Delta_{2n} \\[0.2cm]
		\vdots & {}                & \vdots         &  \vdots    & \vdots  {} &  {} &\vdots \\[0.2cm]
		\Delta_{n1} & \ldots & \Delta_{n,j-1} & \det Df_n (P) & \Delta_{n,j+1} & \ldots & \Delta_{nn} \\[0.2cm]  \end{vmatrix} t
\end{equation}
and
\begin{equation}\label{3.57-A}
	\det(\Delta_j^{(2)})=\begin{vmatrix}
		\Delta_{11} & \ldots & \Delta_{1,j-1} & C_1 & \Delta_{1,j+1} & \ldots & \Delta_{1n} \\[0.2cm]
		\Delta_{21} & \ldots & \Delta_{2,j-1} & C_2 & \Delta_{2,j+1} & \ldots & \Delta_{2n} \\[0.2cm]
		\vdots      & {}     &  \vdots        & \vdots & \vdots & \ldots & \vdots \\[0.2cm]
		\Delta_{n1} & \ldots & \Delta_{n,j-1} & C_n & \Delta_{n,j+1} & \ldots & \Delta_{nn}
	\end{vmatrix}.
\end{equation}
\smallskip
Using the Laplace expansion along the first row, we find that
\begin{equation}\label{3.58}
	\det Df_i(P)
	=\sum_{j=1}^n \frac{\p H}{\p P_j} (-1)^{j+1} \det DF(P)_{j,i} = \sum_{j=1}^n \Delta_{ij} \frac{\p H}{\p P_j}.
\end{equation}
Substituting Eq. \eqref{3.58} into Eq. \eqref{3.56} we obtain
\begin{equation}\label{3.59}
	\det (\Delta_j) = \sum_{k=1}^n \frac{\p H}{\p P_k}
	\begin{vmatrix}
		\Delta_{11} & \ldots & \Delta_{1,j-1} & \Delta_{1k} & \Delta_{1,j+1} & \ldots & \Delta_{1n} \\[0.2cm]
		\Delta_{21} & \ldots & \Delta_{2,j-1} & \Delta_{2k} & \Delta_{2,j+1} & \ldots & \Delta_{2n} \\[0.2cm]
		\vdots      & {}     &  \vdots        & \vdots & \vdots & \ldots & \vdots \\[0.2cm]
		\Delta_{n1} & \ldots & \Delta_{n,j-1} & \Delta_{nk} & \Delta_{n,j+1} & \ldots & \Delta_{nn}
	\end{vmatrix} t + \det(\Delta_j^{(2)}).
\end{equation}
Since all determinants in Eq. \eqref{3.59} multiplying $\frac{\p H}{\p P_k}$ vanish except for $k=j$, the above expression simplifies to
\begin{equation}\label{3.60-A}
	\det (\Delta_j) = \det (\Delta) \frac{\p H}{\p P_j} t + \det (\Delta_j^{(2)}).
\end{equation}
Substituting this into Eq. \eqref{3.54} we find that
\begin{equation}
	Q_j(t) = \frac{\p H}{\p P_j} t + \frac{\det (\Delta_j^{(2)})}{\det(\Delta)}.
\end{equation}
It is easily seen that for given initial conditions $Q_j(0)=Q_{j0}$ we have $\det (\Delta_j^{(2)})=\det (\Delta)\, Q_{j0}$, hence
\begin{equation}\label{3.63-A}
	Q_j(t) = \frac{\p H}{\p P_j} t + Q_{j0}.
\end{equation}
Therefore, along the integral curves of the vector field $A=\p_t + X_H$, the momenta $P_j$ remain constant and the coordinates $Q_j$ linearly evolve in time with frequencies $\frac{\p H}{\p P_j}$. This reproduces the linear flow on invariant tori described by the Arnold--Liouville theorem directly from integration of the Pfaffian forms \eqref{3.47}--\eqref{3.48}. 

We close the section by outlining a practical procedure for reconstructing the action--angle variables based solely on the Pfaffian forms associated with the canonical solvable structure. The action--angle coordinates are recovered locally by successive integration of these Pfaffian forms directly in the physical variables $(q,p)$. Let $\varphi (q,p)=(Q,P)$ denote a symplectomorphism from the physical variables $(q,p)$ to the action--angle variables $(Q,P)$. Let $\omega_i^\prime$ denote the Pfaffian forms
\eqref{3.47}--\eqref{3.48} calculated in the $(q,p)$ variables. Since the pull--back by $\varphi$ preserves the structure of the forms, we have  
\begin{align}
\omega_{n+k}^\prime &= \varphi^\ast \omega_{n+k} = (-1)^{n+k+1}\, d\varphi_{n+k}, \label{3.60-F} \\
\omega_k^\prime &= \varphi^\ast \omega_k = \frac{(-1)^n}{\varphi^\ast \lambda} \Big(\varphi^\ast \det Df_k (P) \, dt+\sum_{j=1}^n (-1)^{j+1} \varphi^\ast \det DF(P)_{j,k}\, dQ_j\Big).  \label{3.61-F}
\end{align}
Now, $Q_i$ and $P_i$ are explicitly recovered using the following procedure:
\begin{enumerate}
	\item The forms $\omega_{n+k}^\prime$ are exact differentials which immediately yields the action variables $P_k=\varphi_{n+k}(q,p)$. 
	\item Having obtained $P_k$, write the Hamiltonian $H$ and the first integrals $F_1,\ldots, F_n$ as functions of $P_1,\ldots, P_n$ to evaluate the Jacobian determinants in the Pfaffian forms \eqref{3.48}.
	\item Now, compare $\omega_k$ with $\omega_k^\prime$
	to obtain the angle variables $Q_k=\varphi_k(q,p)$. 
\end{enumerate}
This procedure is illustrated in the following section by considering the open Toda lattice and the rational Calogero--Moser system.


\section{Examples}

In this section we consider solvable structures for two classical integrable
systems: the open Toda lattice and the rational Calogero--Moser system. Both systems
exhibit scattering--like behaviour since, as $t\to \pm \infty$, the particles separate and 
behave asymptotically as free particles. Although the invariant level sets of the systems are
noncompact, the construction developed in
Section \ref{Sect.-3.1} is local in nature and can be
applied on suitable open subsets of the phase space. This allows us to derive
canonical variables of action--angle type and to integrate the equations of
motion of the systems explicitly.

\subsection{The open Toda lattice}

The Toda lattice was introduced in 1967 as a one--dimensional chain of $n$ particles with nearest--neighbour exponential interaction, originally proposed as a model for wave propagation along a line \cite{Toda1967,Toda1981}. Complete integrability of the finite periodic Toda lattice was established in 1974 in the work of H\'enon, Flaschka, and Manakov \cite{Henon1974,Flaschka1974,Manakov1974}. Flaschka showed that the equations of motion can be written in Lax form $\dot L = [A,L]$ for a pair of $n\times n$ matrices, which implies the eigenvalues of $L$ are constants of motion. Soon afterwards, Moser \cite{Moser1975} showed that the finite nonperiodic (open) Toda lattice is also completely integrable. In Moser's approach, the time evolution of the system is recovered from spectral data of the Lax matrix $L$ consisting of the eigenvalues of $L$ and the spectral weights determined by the Weyl function associated with $L$. In the context of the present work, the spectral formulation of the open Toda lattice is particularly useful because it provides a natural setting for construction of the solvable structure and the corresponding canonical variables of action--angle type in terms of the spectral data. 

The Toda lattice is a system of $n$ particles on a line defined by the Hamiltonian
\begin{equation}\label{115}
H=\frac12 \sum_{k=1}^n p_k^2 + \sum_{k=1}^{n-1} e^{-(q_{k+1}-q_k)}.
\end{equation}
The equations of motion are given by 
\begin{equation}\label{118}
\dot q_k  = p_k,  \quad 
\dot p_k = e^{-(q_{k}-q_{k-1})}-e^{-(q_{k+1}-q_{k})}, \qquad k=1, \ldots, n,
\end{equation}
where, in the nonperiodic case, we set the formal boundary conditions $q_0=-\infty$ and $q_{n+1}=+\infty$ implying $e^{-(q_1-q_0)}=0$ and $e^{-(q_{n+1}-q_n)}=0$. Flaschka \cite{Flaschka1974} introduced the variables
\begin{alignat}{2}
a_k &= e^{-\frac12 (q_{k+1}-q_k)}, & \qquad k &=1, \ldots, n-1, \\
b_k &= -p_k, & \qquad k &= 1, \ldots, n,
\end{alignat}
and showed that Hamilton's equations \eqref{118} are equivalent to the Lax equation
$\dot L = [M,L]$ for the matrices
\begin{equation}\label{LA}
L=\begin{pmatrix}
b_1 & a_1 & 0   & \cdots & 0 \\
a_1 & b_2 & a_2 & \ddots & \vdots \\
0   & a_2 & b_3 & \ddots & 0 \\
\vdots & \ddots & \ddots & \ddots & a_{n-1}\\
0 & \cdots & 0 & a_{n-1} & b_n
\end{pmatrix},
\quad 
M=\begin{pmatrix}
0 & \frac{1}{2} a_1 & 0   & \cdots & 0 \\
-\frac{1}{2} a_1 & 0 & \frac{1}{2} a_2 & \ddots & \vdots \\
0   & -\frac{1}{2} a_2 & 0 & \ddots & 0 \\
\vdots & \ddots & \ddots & \ddots & \frac{1}{2} a_{n-1} \\
0 & \cdots & 0 & -\frac{1}{2}a_{n-1} & 0
\end{pmatrix}.
\end{equation}
Since $L$ is a real Jacobi matrix, it has $n$ distinct eigenvalues which we arrange as $\lambda_1< \lambda_2< \ldots< \lambda_n$. The eigenvalues of $L$ or, equivalently, the functions $F_k=\Tr (L^k)$ provide $n$ independent integrals of the system in involution. 

In the following, we construct the canonical solvable structure given by Theorem \ref{tm-3.2} for the distribution $\D=\Span \{A\}$ where $A=\p_t + X_H$ and 
\begin{equation}
X_H = \sum_{k=1}^n p_k \frac{\p}{\p q_k} + \sum_{k=1}^n \Big(e^{-(q_{k}-q_{k-1})}-e^{-(q_{k+1}-q_{k})}\Big)\frac{\p}{\p p_k}
\end{equation}
We will show that if the first integrals are given by $F_k=\Tr (L^k)$, then the functions $G_k$ can be defined in terms of the spectral variables discovered by Vaninsky \cite{Vaninsky2002} and expressed as functions on the reduced phase space. From this perspective, Vaninsky's variables generate the Hamiltonian vector fields $X_{G_k}$ required in the solvable structure. 

The Hamiltonian \eqref{115} is invariant under translations, hence the total momentum $P_c = \sum_{k=1}^n p_k$ is conserved along the flow of $X_H$. Thus, for a fixed value of $P_c$, the dynamics of the system can be completely described in terms of the reduced phase space variables 
\begin{equation}\label{122}
x_k = q_{k+1}-q_k, \quad y_k = \sum_{i=k+1}^n p_i - \frac{n-k}{n} P_c, \qquad k=1, \ldots, n-1.
\end{equation}
Here, $x_k$ are the relative positions and $y_k$ are the conjugate momenta satisfying the canonical Poisson bracket $\{x_i,x_j\}=\{y_i,y_j\}=0$ and $\{x_i,y_j\}=\delta_{ij}$. The reduced phase space, denoted $M_{red}$, is a symplectic manifold with the symplectic form
\begin{equation}
	\omega_{red} = \sum_{i=1}^{n-1} dx_i \wedge dy_i.
\end{equation}
The dimension of $M_{red}$ is $\dim M_{red}=2n-2$ which
allows the rank of the solvable structure to be reduced by two. Since $F_1 = \operatorname{tr}(L)=-P_c$, it is sufficient to use the $2n-2$ functions $F_i$ and $G_i$, $\, 2\leq i \leq n$, to construct the solvable structure for $\D=\Span \{A\}$. 

The key object in defining the Vaninsky's spectral variables is the Weyl function associated with $L$,
\begin{equation}
w(\lambda) = \langle (L-\lambda I)^{-1} e_1, e_1\rangle
\end{equation}
where $\langle \, , \, \rangle$ denotes the inner product on $\R^n$ and $e_1 = (1, 0, \ldots, 0)$ is the first vector of the canonical basis of $\R^n$. Using the spectral decomposition $L=\sum_{k=1}^n \lambda_k\, P_k$ where $P_k = \langle \cdot \, , v^{(k)}\rangle\, v^{(k)}$ is the projection on the normalized eigenvector $v^{(k)}$ of $L$ corresponding to $\lambda_k$, one can show that 
\begin{equation}\label{124}
w(\lambda) = \sum_{k=1}^n \frac{r_k}{\lambda_k - \lambda}, \quad r_k = (v_1^{(k)})^2,
\end{equation}
where $v_1^{(k)}$ denotes the first component of $v^{(k)}$. Thus, the function $w(\lambda)$ has simple poles at $\lambda=\lambda_k$ with residues $r_k$, called Moser's spectral weights. Since $L$ is a Jacobi matrix, the spectral weights satisfy the constraint $\sum_{k=1}^n r_k=1$. Moser \cite{Moser1975} showed, using the Weyl function, that if $L$ satisfies the Lax equation $\dot L = [M,L]$, then the spectral weights $r_k$ are rational functions of the exponentials $\exp(\lambda_j t)$. The spectral data $\{\lambda_k, r_k\}$ uniquely determine the matrix $L$, and thus provide a complete description of the Toda flow. Along the Toda flow, the eigenvalues of $L$ satisfy the constraint $\sum_{k=1}^n \lambda_k =-P_c$. Hence, only $n-1$ eigenvalues are independent, which we choose to be $\lambda_2, \ldots, \lambda_n$. While the constraint on the eigenvalues $\lambda_k$ is a consequence of the time evolution of $L$, the constraint on the spectral weights $r_k$ holds for any Jacobi matrix.

For the purpose of the present work, it is convenient to represent the Weyl function as the ratio $w(\lambda)=-\frac{q(\lambda)}{p(\lambda)}$ where $p(\lambda) = \prod_{k=1}^n (\lambda_k - \lambda)$ and $q(\lambda)$ is the monic polynomial of degree $n-1$ given by
\begin{equation}
q(\lambda) = -\sum_{k=1}^n \frac{p(\lambda)}{\lambda_k - \lambda}\, r_k.
\end{equation}
In studying the Atiyah--Hitchin bracket for the Weyl function \eqref{124}, Vaninsky \cite{Vaninsky2002} showed that if we take the independent eigenvalues $\lambda_2, \ldots, \lambda_n$ and define the variables 
\begin{equation}
\theta_k = \ln \left((-1)^{k+1} \frac{q(\lambda_k)}{q(\lambda_1)}\right), \qquad k=2, \ldots, n,
\end{equation}
then $\lambda_k$ and $\theta_k$ satisfy the canonical Poisson bracket
\begin{equation}
\{\lambda_i,\lambda_j\} = \{\theta_i,\theta_j\}=0, \quad \{\theta_i,\lambda_j\}=\delta_{ij}, \quad i,j=2, \ldots, n.
\end{equation}
(We remark that in Ref. \cite{Vaninsky2002}, the independent variables are labeled by $\lambda_1, \ldots, \lambda_{n-1}$ and $\theta_1, \ldots, \theta_{n-1}$.) The variables $(\theta,\lambda)$, called spectral variables, can be expressed in terms of the reduced phase space variables \eqref{122}, and the map $(x,y)\mapsto (\theta,\lambda)$ preserves the symplectic form on $M_{red}$ so that
\begin{equation}
\omega_{red} = \sum_{i=2}^n d\theta_i \wedge d\lambda_i.
\end{equation}
Therefore, $(\theta,\lambda)$ can be taken as the Darboux coordinates on $M_{red}$. 

Consider now the Toda lattice with fixed total momentum $P_c=-\sum_{i=1}^n \lambda_i$ and define the map 
$F\colon (\lambda_2,\ldots, \lambda_n)\mapsto (F_2,\ldots, F_n)$ where $F_k=\Tr (L^k) = \sum_{i=1}^n \lambda_i^k$. Since the dependent eigenvalue is given by $\lambda_1=-\sum_{i=2}^n \lambda_i - P_c$, we have 
\begin{equation}
	\frac{\p F_k}{\p \lambda_j} = k (\lambda_j^{k-1}-\lambda_1^{k-1}), \quad j,k=2, \ldots, n.
\end{equation}
Therefore, the Jacobian of $F$ is given by
\begin{equation}
	DF(\lambda)=\begin{bmatrix} 2(\lambda_2-\lambda_1) & 2(\lambda_3-\lambda_1) & \ldots & 2 (\lambda_n-\lambda_1) \\
		3(\lambda_2^2-\lambda_1^2) & 3(\lambda_3^2-\lambda_1^2) & \ldots & 3 (\lambda_n^2-\lambda_1^2) \\
		\vdots                      & {}                         & {}     & \vdots                        \\
		\vdots                      & {}                         & {}     & \vdots                        \\
		n (\lambda_2^{n-1} - \lambda_1^{n-1}) & n (\lambda_3^{n-1}-\lambda_1^{n-1}) & \ldots & n (\lambda_n^{n-1}-\lambda_1^{n-1})
	\end{bmatrix}
\end{equation}
and the determinant of $DF(\lambda)$ has the Vandermonde form 
\begin{equation}
\det DF(\lambda) = n!\, \prod_{i<j} (\lambda_j-\lambda_i). 
\end{equation}
Since $\lambda_i \neq \lambda_j$ for $i\neq j$, the map $F$ is locally invertible. Thus, by Theorem \ref{tm-3.2}, the functions $G_k$ are locally given by Eq. \eqref{22-A}, 
\begin{equation}\label{134-A}
	G=\frac{1}{n!\, \prod_{i<j} (\lambda_j-\lambda_i)} \operatorname{cof}(DF(\lambda))\, \theta
\end{equation}
where $\operatorname{cof}(DF(\lambda))$ is the cofactor matrix of $DF(\lambda)$. 
This completely determines the canonical solvable structure $(A,X_{F_2}, \ldots, X_{F_n}, X_{G_2}, \ldots, X_{G_n})$ for the open Toda lattice in the spectral variables $(\theta,\lambda)$ with first integrals $F_k = \sum_{i=1}^n \lambda_i^k$. 

Explicit integration of the open Toda lattice has been considered by several authors \cite{OP1979, Ko1979}. To illustrate the proposed method, we derive explicit solutions of the system with the Hamiltonian $H=\frac12 (p_1^2+p_2^2)+e^{-(q_2-q_1)}$. 
The Lax matrix is given by 
\begin{equation}
L=\begin{bmatrix} b_1 & a_1 \\ a_1 & b_2\end{bmatrix}, \qquad a_1=e^{-\frac12 (q_2-q_1)}, \quad b_i=-p_i,
\end{equation}
which yields the associated Weyl function
\begin{equation}
w(\lambda) = \langle (L-\lambda I)^{-1} e_1, e_1\rangle = \frac{b_2-\lambda}{\det (L-\lambda I)}.
\end{equation}
This implies that the polynomial $q(\lambda)$ is given by $q(\lambda)=\lambda-b_2$.   
The eigenvalues of $L$ arranged as $\lambda_1 < \lambda_2$ are found to be 
\begin{equation}
	\lambda_1 = \frac12 (b_1+b_2-\sqrt{\Delta}), \quad \lambda_2 = \frac12 (b_1+b_2+\sqrt{\Delta})
\end{equation}
where $\Delta = (b_1-b_2)^2 + 4 a_1^2$. The spectral variables are the pair $(\theta_2, \lambda_2)$ where 
\begin{equation}
\theta_2 = \ln\left(-\frac{q(\lambda_2)}{q(\lambda_1)}\right).
\end{equation}
Using the reduced phase space variables $x=q_2-q_1$ and $y=p_2-\frac12 P_c$ where $P_c=p_1+p_2$, we find
\begin{equation}
\theta_2 = \ln \left(\frac{\sqrt{y^2+e^{-x}}+y}{\sqrt{y^2+e^{-x}}-y}\right), \quad \lambda_2=-\frac12 P_c + \sqrt{y^2+e^{-x}}.
\end{equation}

We next determine the canonical solvable structure $(A,X_{F_2},X_{G_2})$ for $A=\p_t+X_H$. In the spectral variables we have $H=\frac12 (\lambda_1^2+\lambda_2^2)$ and $F_2= \lambda_1^2+\lambda_2^2$. It follows from Eq. \eqref{134-A} that $G_2$ is given by
\begin{equation}
	G_2 = \frac{1}{2 (\lambda_2-\lambda_1)} \theta_2.
\end{equation}
This yields the solvable structure
\begin{align}
	A &= \p_t + (\lambda_2-\lambda_1) \frac{\p}{\p \theta_2}, \\
	X_{F_2} &= 2 (\lambda_2-\lambda_1) \frac{\p}{\p \theta_2}, \\
	X_{G_2} &= -\frac{\theta_2}{(\lambda_2-\lambda_1)^2} \frac{\p}{\p \theta_2}-
	\frac{1}{2 (\lambda_2-\lambda_1)}\frac{\p}{\p \lambda_2}.
\end{align}
Introduce the volume form $\tau = dt \wedge d\theta_2 \wedge d\lambda_2$. Then the integrating factor is given by $\lambda = X_{G_2}\, \lrcorner \, X_{F_2}\, \lrcorner \, A \, \lrcorner\, \tau = -1$, and the associated Pfaffian forms are found to be 
\begin{align}
	\omega_2 &= \frac{1}{\lambda} X_{G_2} \, \lrcorner \, A \, \lrcorner \, \tau = -dF_2, \label{144} \\
	\omega_1 &= \frac{1}{\lambda} X_{F_2}\, \lrcorner \, A \, \lrcorner \, \tau = \frac12 dt - dG_2. \label{145}
\end{align}
Along the integral curves $\gamma$ of the vector field $A$ we have $\omega_1\vert_\gamma = \omega_2\vert_\gamma =0$ which implies
\begin{align}
F_2 &= \lambda_1^2+\lambda_2^2 = C_2, \label{146} \\
G_2 &= \frac{1}{2(\lambda_2-\lambda_1)} \theta_2 = \frac12 t + C_1 \label{147}
\end{align}
for some $C_1,C_2\in \R$. Expressing $F_2$ in the reduced variables we find from \eqref{146} that
\begin{equation}\label{148} 
y^2 + e^{-x} = K_1^2, \quad K_1^2 = \frac12 C_2 - \frac{1}{4} P_c^2.
\end{equation}
This implies that on the integral curve $\gamma$ the spectral variable $\theta_2$ is given by $\theta_2=\ln\left(\frac{K_1+y}{K_1-y}\right)$. Hence, it follows from \eqref{147} that 
\begin{equation}\label{149}
\ln\left(\frac{K_1+y}{K_1-y}\right) = 2K_1 t + 2K_2, \quad K_2 = 2K_1 C_1.
\end{equation}
The explicit solution of the system \eqref{148}--\eqref{149} is found to be 
\begin{align}
x &=2 \ln\Big(\frac{1}{K_1} \cosh (K_1 t + K_2)\Big), \\
y &= K_1 \tanh (K_1 t + K_2).
\end{align}
Recovering the phase space variables $q_i,p_i$ from $x$ and $y$ for a fixed value of the 
total momentum $P_c$, we find
\begin{align}
q_1(t) &= \frac12 P_c\ t - \ln\Big(\frac{1}{K_1} \cosh(K_1 t+K_2)\Big)+Q_c(0), \\
q_2(t) &= \frac12 P_c\, t + \ln\Big(\frac{1}{K_1} \cosh(K_1 t+K_2)\Big)+Q_c(0), \\
p_1(t) &= \frac12 P_c - K_1 \tanh (K_1 t+K_2), \\
p_2(t) &= \frac12 P_c + K_1 \tanh (K_1 t+K_2),
\end{align}
where $Q_c(0)$ is the initial position of the center of mass $Q_c = \frac12 (q_1+q_2)$. We note that the particles behave asymptotically as free particles since 
\begin{equation}
	q_i(t) \sim v_i^{\pm} t + a_i^{\pm} \quad \text{as}\quad t\to \pm \infty
\end{equation}
where $v_i^{\pm}$ are the asymptotic velocities
\begin{equation}
	v_1^+ = v_2^- = \frac{1}{2}P_c - K_1, \quad v_2^+ = v_1^- = \frac{1}{2} P_c+K_1.
\end{equation}
Thus, as $t$ evolves from $-\infty$ to $+\infty$, the particles exchange their asymptotic velocities. 

The Pfaffian forms $\omega_1$ and $\omega_2$ yield canonical variables of action--angle type for the Toda system. Comparing Eqs. \eqref{144} and \eqref{145} with the general expressions \eqref{3.47} and \eqref{3.48}, we conclude that $Q=G_2$ and $P=F_2$. Expressing the eigenvalues $\lambda_1$ and $\lambda_2$ in terms of the reduced phase space variables, we find that 
\begin{equation}
	Q=\frac{1}{4 \sqrt{y^2+e^{-x}}} \ln\left(\frac{\sqrt{y^2+e^{-x}}+y}{\sqrt{y^2+e^{-x}}-y}\right), \quad P=\frac12 P_c^2+2(y^2+e^{-x}).
\end{equation}


\subsection{Rational Calogero--Moser system}

The Calogero--Moser system was first introduced by F. Calogero in the early 1970s \cite{Ca_1971} as a one--dimensional system of $n$ identical particles interacting through the rational pairwise potential $V_{ij}=g^2 /(q_i-q_j)^2$ and, in some variants, through trigonometric and hyperbolic potentials. J. Moser \cite{Mo_1975} established complete integrability of the classical rational Calogero model by showing that the equations of motion are equivalent to the Lax equation $\dot L = [L,M]$ for a pair of $n\times n$ matrices. 
The eigenvalues of $L$ or, equivalently, the functions $F_k=\operatorname{tr}(L^k)$, $k=1,\ldots, n$, provide $n$ independent integrals in involution. Subsequent work by Wojciechowski \cite{Wo_1983} showed that the Calogero--Moser system is in fact superintegrable, possessing $2n-1$ independent first integrals. Later investigations by Ruijsenaars \cite{Ru_1988} and 
Brzezinski et al. \cite{Br_1999} developed different methods for the construction of action--angle variables of the Calogero--Moser systems, although the explicit computations are rather involved. For additional details on the Calogero--Moser systems, see \cite{Ca_2001}.

In this section we consider the rational Calogero--Moser system defined by the Hamiltonian
\begin{equation}\label{3.72}
	H=\frac{1}{2} \sum_{i=1}^n p_i^2 + g^2 \sum_{i<j} \frac{1}{(q_i-q_j)^2}
\end{equation}
where $g^2>0$ is the coupling constant giving rise to a repulsive pairwise potential. The Lax pair \cite{Mo_1975} for the system is given by the $n\times n$ matrices with elements
\begin{equation}
	L_{ij} = p_i\, \delta_{ij}+ \sqrt{-1}\, g\, \frac{1-\delta_{ij}}{q_i - q_j}
\end{equation}
and
\begin{equation}
	M_{ij} = \sqrt{-1}\, g\left(-\delta_{ij} \sum_{k\neq i} \frac{1}{(q_i-q_k)^2}+
	\frac{1-\delta_{ij}}{(q_i-q_j)^2}\right).
\end{equation}
The Lax equation $\dot L = [L,M]$ is equivalent to Hamilton's equations
\begin{equation}
\dot q_i = \frac{\p H}{\p p_i}= p_i, \quad \dot p_i = -\frac{\p H}{\p q_i} = 2g^2 \sum_{j\neq i} \frac{1}{(q_i-q_j)^3}, 
\qquad i=1, \ldots, n.
\end{equation}

To keep the algebra simple, we illustrate the construction of a solvable structure in the two--particle case $n=2$. Then the Hamiltonian vector field is given by
\begin{equation}
	X_H = p_1 \frac{\p}{\p q_1}+p_2 \frac{\p}{\p q_2} + \frac{2g^2}{(q_1-q_2)^3} \Big(\frac{\p}{\p p_1}-\frac{\p}{\p p_2}\Big).
\end{equation}
The conserved quantities $F_k=\mbox{tr}(L^k)$ are found to be
\begin{equation}\label{3.77}
	F_1 = p_1 + p_2, \quad F_2 = p_1^2+p_2^2+\frac{2g^2}{(q_1-q_2)^2}.
\end{equation}
We construct a solvable structure of the form
\begin{align}
	&[X_{G_1},X_{G_2}] =0, \label{3.78} \\
	&[X_{G_i},X_{F_j}] = f_{ij}^1\, X_{F_1}+f_{ij}^2\, X_{F_2}, \label{3.79} \\
	&[X_{G_i},A] = h_{i1}\, X_{F_1} + h_{i2}\, X_{F_2},   \label{3.80}
\end{align}
for the distribution $\D=\Span \{A\}$ with $A=\p_t + X_H$. According to Theorem \ref{tm-3.3}, if $G_1$ and $G_2$ satisfy $\{G_i,F_j\}=K_{ij}(F_1,F_2)$ for some smooth function $K_{ij}$, then
$[X_{G_i},X_{F_j}]\in \operatorname{span} \{X_{F_1},X_{F_2}\}$. This condition leads to the system of equations
\begin{align}
	&\frac{\p G_i}{\p q_1} + \frac{\p G_i}{\p q_2} = K_{i1} (F_1,F_2), \label{3.81}\\
	&2p_1\, \frac{\p G_i}{\p q_1} + 2p_2\, \frac{\p G_i}{\p q_2} + \frac{4g^2}{(q_1-q_2)^3}\Big(\frac{\p G_i}{\p p_1} - \frac{\p G_i}{\p p_2}\Big) = K_{i2}(F_1,F_2), \quad i=1,2.   \label{3.82}
\end{align}
For $i=1$, Eqs. \eqref{3.81}--\eqref{3.82} have a simple solution
\begin{equation}\label{4.87}
	G_1=q_1 + q_2
\end{equation}
with $K_{11}=2$ and $K_{12}=2F_1$. For $i=2$, if we assume that $(G_2)_{q_1}+(G_2)_{q_2}=0$, 
then Eq. \eqref{3.81} implies $K_{21}=0$. Since $G_1$ and $G_2$ are required to Poisson commute, $\{G_1,G_2\}=0$, it follows that $(G_2)_{p_1}+(G_2)_{p_2}=0$. The simplest choice for $G_2$ satisfying both conditions is
\begin{equation}\label{4.88}
	G_2 = (q_1-q_2)(p_1-p_2).
\end{equation}
With this choice for $G_2$, Eq. \eqref{3.82} yields
\begin{equation}
	K_{22} = 4 F_2-2 F_1^2.
\end{equation}
The Hamiltonian vector fields $X_{F_1}, X_{F_2}, X_{G_1}, X_{G_2}$ are linearly independent on the open set $U=\{(q,p)\in \R^4\mid q_1\neq q_2\}$, and the solvable structure is given by
\begin{alignat}{2}
	&[X_{G_1},X_{G_2}] =0, &\qquad {}\\
	&[X_{G_1},X_{F_1}] =0, &\qquad &[X_{G_1},X_{F_2}] =-2 X_{F_1}, \\
	&[X_{G_2},X_{F_1}] =0, &\qquad &[X_{G_2},X_{F_2}] = 4(p_1+p_2) X_{F_1} - 4 X_{F_2}, \\
	&[X_{G_1},A] =-X_{F_1}, &\qquad &[X_{G_2},A] = 2(p_1+p_2) X_{F_1} - 2 X_{F_2}.
\end{alignat}
Note that the above solvable structure is not canonical since $K_{ij}(F_1,F_2)\neq \delta_{ij}$.

We proceed to compute the Pfaffian forms \eqref{3.40}--\eqref{3.41} associated with the solvable structure. Since the Jacobian determinant
\begin{equation}\label{3.86}
	\left\vert \frac{\p (F_1,F_2,G_1,G_2)}{\p (q_1,q_2,p_1,p_2)}\right\vert = 4(p_1-p_2)^2 + \frac{16 g^2}{(q_1-q_2)^2}
\end{equation}
is nonvanishing on $U$, we can express the vector fields $X_{F_i}$, $X_{G_i}$ in terms of the variables $(F_1,F_2,G_1,G_2)$ which simplifies the computation of the Pfaffian forms. We have
\begin{align}
	A &= \frac{\p}{\p t} + F_1\, \frac{\p}{\p G_1} + (-F_1^2+2F_2)\, \frac{\p}{\p G_2}, \\
	X_{F_1} &= 2\, \frac{\p}{\p G_1}, \\
	X_{F_2} &= 2 F_1\, \frac{\p}{\p G_1} + 2(-F_1^2+2 F_2)\, \frac{\p}{\p G_2}, \\
	X_{G_1} &= -2\, \frac{\p}{\p F_1} - 2 F_1\, \frac{\p}{\p F_2}, \\
	X_{G_2} &= -2(-F_1^2+2F_2)\, \frac{\p}{\p F_2},
\end{align}
and the volume form is given by
\begin{equation}
	\tau = \frac{1}{\lambda} dt \wedge dF_1 \wedge dF_2 \wedge dG_1 \wedge dG_2, \quad \lambda = 4(-F_1^2+2F_2).
\end{equation}
By straightforward calculation we find that
\begin{align}
	\omega_4 &= \frac{1}{\lambda} X_{G_1} \lrcorner X_{F_2} \lrcorner X_{F_1} \lrcorner A \lrcorner \tau =
	\frac{1}{2(-F_1^2+2F_2)}\big(F_1 dF_1 - dF_2),  \label{4.w_4} \\[0.2cm]
	\omega_3 &= \frac{1}{\lambda} X_{G_2} \lrcorner X_{F_2} \lrcorner X_{F_1} \lrcorner A \lrcorner \tau =
	\frac{1}{2} dF_1,  \label{4.w_3} \\[0.2cm]
	\omega_2 &= \frac{1}{\lambda} X_{G_2} \lrcorner X_{G_1} \lrcorner X_{F_1} \lrcorner A \lrcorner \tau =
	-\frac{1}{2} dt + \frac{1}{2(-F_1^2+2F_2)} dG_2, \label{4.w_2} \\[0.2cm]
	\omega_1 &= \frac{1}{\lambda} X_{G_2} \lrcorner X_{G_1} \lrcorner X_{F_2} \lrcorner A \lrcorner \tau =
	-\frac{1}{2} dG_1 + \frac{F_1}{2(-F_1^2+2F_2)} dG_2.  \label{4.w_1}
\end{align}
The last two forms are exact differentials,
\begin{alignat}{2}
	\omega_4 &= dI_4, &\qquad I_4 &= -\frac{1}{4} \ln\big(-F_1^2+2F_2\big), \label{3.101}\\
	\omega_3 &= dI_3, &\qquad I_3 &= \frac{1}{2} F_1,   \label{3.102}
\end{alignat}
in accordance with Eq. \eqref{3.47}. 
Let $\gamma$ be an integral curve of $A=\p_t+X_H$. Then $\omega_3\vert_\gamma = \omega_4
\vert_\gamma = 0$, hence $I_3$ and $I_4$ are first integrals of $A$. 
Define the submanifold
\begin{equation}
	M_2=\Big\{(t,q,p)\mid I_3=c_3, \; I_4=c_4\Big\}.
\end{equation}
Then the restriction of $\omega_2$ to $M_2$ is exact,
\begin{equation}
	\omega_2\vert_{M_2} = dI_2, \quad I_2 = -\frac{1}{2} t + \frac{1}{2} e^{4c_4}\, G_2.
\end{equation}
On the integral curve $\gamma$ we have $\omega_2\vert_{M_2}=0$, hence $I_2$ is a first integral of $A$.
Finally, the lowest form $\omega_1$ is exact on the submanifold
\begin{equation}
	M_1 = \Big\{(t,q,p) \mid I_2=c_2, \; I_3=c_3, \; I_4=c_4\Big\}
\end{equation}
and
\begin{equation}
	\omega_1\vert_{M_1} = dI_1, \qquad I_1=c_3 t - \frac{1}{2} G_1.
\end{equation}
Since $\omega_1\vert_{M_1}$ vanishes on $\gamma$, it follows that $I_1$ is a first
integral of $A$.
Therefore, $\gamma$ is the submanifold $\gamma=\{(t,q,p)\mid I_i=c_i,\; 1\leq i \leq 4 \}$.
Expressing the functions $F_i$ and $G_i$ in terms of the integrals $I_i$, we find that $\gamma$
is defined by the system of equations
\begin{align}
	p_1 + p_2 &= 2 c_3, \\
	p_1^2 + p_2^2 + \frac{2 g^2}{(q_1-q_2)^2} &= \frac{1}{2} e^{-4c_4} + 2 c_3^2, \\
	q_1+q_2 &= 2 (c_3 t - c_1), \\
	(q_1-q_2)(p_1-p_2) &= e^{-4c_4} (t+2c_2)
\end{align}
which represents the solution of the Calogero--Moser system in implicit form. 
Solving the above system for $q_i$ and $p_i$, we obtain the explicit form of the solution
\begin{align}
	q_1(t) &= c_3 t - c_1 +\frac{1}{2} \Big[e^{-4 c_4}(t+2c_2)^2 + 4g^2 e^{4 c_4}\Big]^{\frac12}, \\
	q_2(t) &= c_3 t - c_1 -\frac{1}{2} \Big[e^{-4 c_4}(t+2c_2)^2 + 4g^2 e^{4 c_4}\Big]^{\frac12}, \\
	p_1(t) &= c_3 + \frac{1}{2} \frac{e^{-4 c_4} (t+2 c_2)}{\Big[e^{-4 c_4}(t+2 c_2)^2+4g^2 e^{4 c_4}\Big]^\frac12}, \\
	p_2(t) &= c_3 - \frac{1}{2} \frac{e^{-4 c_4} (t+2 c_2)}{\Big[e^{-4 c_4}(t+2 c_2)^2+4g^2 e^{4 c_4}\Big]^\frac12},
\end{align}
where the constants $c_i$ are determined from the initial conditions $q_i(0)=q_{i0}$, $p_i(0)=p_{i0}$. The asymptotic behaviour of the particles is given by 
\begin{equation}
	q_i(t) \sim v_i^{\pm} t + a_i^{\pm} \quad \text{as}\quad t\to \pm \infty
\end{equation}
where $v_i^{\pm}$ are the asymptotic velocities 
\begin{equation}
	v_1^+=v_2^- = c_3 +\frac{1}{2} e^{-2 c_4}, \quad v_2^+ = v_1^- = c_3-\frac{1}{2} e^{-2c_4}.
\end{equation}
Thus, the particles behave asymptotically as free particles and exchange their velocities as $t$ evolves
from $-\infty$ to $+\infty$. 

Next, we determine canonical variables of action--angle type for the Calogero--Moser system. 
Comparing Eq. \eqref{3.47} with \eqref{3.101}--\eqref{3.102}, we find that the action variables are given by
\begin{equation}\label{4.P}
	P_1 = \frac{1}{2}(p_1+p_2), \quad P_2 = \frac{1}{4} \ln \left((p_1-p_2)^2 + \frac{4g^2}{(q_1-q_2)^2}\right).
\end{equation}
The Hamiltonian $H$
and first integrals $F_k$ can be expressed as functions of $P_1$ and $P_2$ as 
\begin{equation}\label{4.123}
	H=P_1^2+\frac{1}{4} e^{4 P_2}, \quad F_1=2 P_1, \quad F_2 = 2 P_1^2 + \frac{1}{2} e^{4 P_2}.
\end{equation}
Now, the Pfaffian forms \eqref{3.48} are found to be
\begin{align}
	\omega_1 &= \frac{1}{\lambda} \left( \Big\vert \frac{\p (H,F_2)}{\p (P_1,P_2)}\Big\vert dt - \frac{\p F_2}{\p P_2} dQ_1
	+\frac{\p F_2}{\p P_1} dQ_2\right) = -\frac{1}{2} dQ_1 + \frac{P_1}{e^{4P_2}} dQ_2, \label{4.124}  \\
	\omega_2 &= \frac{1}{\lambda} \left(\Big\vert \frac{\p (H,F_1)}{\p (P_1,P_2)}\Big\vert dt - \frac{\p F_1}{\p P_2} dQ_1+
	\frac{\p F_1}{\p P_1} dQ_2\right) = -\frac{1}{2} dt + \frac{1}{2 e^{4P_2}} dQ_2.    \label{4.125}
\end{align}
If we substitute Eq. \eqref{4.123} for $F_1$ and $F_2$ into \eqref{4.w_2}--\eqref{4.w_1}, then a comparison with
\eqref{4.124}--\eqref{4.125} yields the angle variables
\begin{equation}\label{4.Q}
	Q_1 = G_1 = q_1+q_2, \quad Q_2 = G_2 = (q_1-q_2)(p_1-p_2).
\end{equation}
One easily verifies that $(q,p)\mapsto (Q,P)$ is a symplectic transformation, hence the variables $(Q,P)$ satisfy the canonical Poisson bracket, as required.

\section{Conclusion}

In this paper we have shown that solvable structures provide a natural geometric
framework for the integration of completely integrable Hamiltonian systems. Starting
from a family of independent first integrals in involution, we constructed a canonical
solvable structure on the extended phase space and showed how the associated Pfaffian
forms encode both the integration of Hamilton's equations and the construction of
canonical variables of action--angle type. In this way, the Arnold--Liouville theorem
is reinterpreted from the point of view of solvable structures: the integration of the
Hamiltonian flow is reduced to the successive integration of a specific system of
Pfaffian forms. The examples of the open Toda lattice and the rational Calogero--Moser
system illustrate that the method is not merely formal, but can be used to recover
explicit solutions and canonical variables in concrete integrable models.

The present approach suggests several directions for future investigations.
One natural question is whether the
construction can be extended to noncommutative integrable systems, where the constants
of motion do not necessarily Poisson commute and the geometry is governed by
nonabelian symmetry algebras and moment maps. Another natural direction concerns applications of \(C^\infty\)-structures to Hamiltonian systems, as they provide a generalization of solvable
structures. It would also be interesting
to investigate dissipative systems within this framework, especially those admitting
contact geometric formulations. Finally, solvable
structures may be useful in systems which do not possess a sufficient number of known
first integrals, since they can sometimes reveal hidden integration mechanisms beyond
the standard Liouville scheme.  These questions suggest that solvable structures may serve as a bridge between classical Hamiltonian integrability and broader geometric approaches to the
study of dynamical systems.

\section*{Funding}

This research was funded by the European Union (NextGenerationEU) under the Croatian Recovery and Resilience Plan 2021–2026 (NRRP), through the University of Split institutional project “Mathematical modeling and simulations of physical systems (MaMoS) IP-UNIST-46”, approved by the Ministry of Science, Education and Youth of the Republic of Croatia.\\
C. Muriel and A. Ruiz also acknowledge the support from {\it   Junta de Andaluc\'ia} (Spain)  to the  research group FQM--377.

\bibliographystyle{unsrt}
\bibliography{references}
\end{document}